\documentclass{aastex701}
\usepackage{upgreek}

\begin{document}

\title{Trial dispersion measure spacing in fast radio burst searches with HEIMDALL} 


\author[0000-0002-4553-655X]{E. F. Keane}
\affiliation{School of Physics, Trinity College Dublin, College Green, Dublin 2, D02 PN40, Ireland}
\email{evan.keane@tcd.ie}

\author[0000-0001-7185-1310]{D. J. McKenna}
\affiliation{ASTRON, The Netherlands Institute for Radio Astronomy, Oude Hoogeveensedijk 4, 7991 PD Dwingeloo, The Netherlands}
\email{mckenna@astron.nl}

\begin{abstract} 
Fast radio bursts (FRBs) are brief flashes of emission detectable to
cosmological distances. Cosmology applications rely on an
understanding of how the detected sample relates to the underlying
population. To this end we examine the dispersion measure `tolerance'
parameter employed by the FRB search tool \textsc{heimdall} and
provide the relation between this and minimum search depth.  Several
FRB samples can be `retro-fitted' using this to more properly account
for survey completeness.
\end{abstract}

\keywords{surveys --- radio astronomy --- astronomy data analysis --- fast radio bursts}

\section{HEIMDALL \& DEDISP}
Many samples of fast radio bursts (FRBs) have been discovered using
the data analysis package \textsc{heimdall}~\citep{heimdall}. This
software analyses \textsc{sigproc}~\citep{sigproc} filterbank data,
i.e.  time-frequency-Stokes~$I$ data cubes. It performs a search in
dispersion measure (DM) across a range of values, then
boxcar-threshold searches the resultant time series. This has been
employed in new discoveries and studies of FRB
sources~\citep{tsb+13,SUPERB2,bsh+25} and
pulsars~\citep{12secPSR,1913fast}, as well as studies of the
time-variable nature of the local radio spectrum at
observatories~\citep{kblp26}. In order to perform dedispersion
\textsc{heimdall} invokes the \textsc{dedisp} package~\citep{dedisp},
GPU-based dedispersion software. One can specify many parameters for
the search, e.g. the threshold signal-to-noise (S/N) ratio, the DM
search range, various parameters for clustering multiple detections of
the same event. In this research note we discuss the DM tolerance
parameter, `dm\_tol'. This parameter can be used to specify the
maximum acceptable S/N drop experienced for pulses with DM falling
`between' search trial values. 

\section{DM tolerance}
The value of $\mathrm{dm\_tol}$ is greater than $1$. Its meaning is
often misinterpreted so that the default value of $1.25$ is taken to
mean: (i) $100\times(\mathrm{dm\_tol}-1)\%=25\%$ S/N loss between DM
trials; or (ii) a $100\times(1-1/(\mathrm{dm\_tol})\%=20\%$ loss in
S/N between DM trials. It is neither of these. It is defined by
\citet{lina_thesis} so that the effective smeared pulse duration
(hereafter effective width) at the $(i+1)^{\rm th}$ DM trial is
$\mathrm{dm\_tol}$ times the effective width at the $(i)^{\rm th}$ DM
trial, i.e. $W_{\rm eff,i+1}=\mathrm{dm\_tol} \times W_{\rm eff,i}$.
The effective width is given by: $W_{\rm eff} = \left(t_{\rm int}^2 +
t_{\rm samp}^2 + t_{\rm DM}^2 + t_{\updelta\mathrm{DM}}^2 + \tau_{\rm
s}^2\right)^{1/2}$, where in \textsc{dedisp} the intrinsic width
$t_{\rm int}$ is hard-coded to $40\;\upmu\mathrm{s}$, $t_{\rm samp}$
is the sampling time, $t_{\rm DM}$ is the dispersion smearing within a
frequency channel, $t_{\rm \updelta\mathrm{DM}}$ is the disperion
smearing across the band due to the DM being off by
$\updelta\mathrm{DM}$ from the true DM, and $\tau_{\rm s}$ is the
scattering time. As the S/N scales as the square root of the effective
width, the meaning of $\mathrm{dm\_tol}$ is also sometimes interpreted
to mean: (iii) $100\times(\sqrt{\mathrm{dm\_tol}}-1)\%$ loss in S/N
between DM trials, i.e. $\sim 11.6\%$ loss for
$\mathrm{dm\_tol}=1.25$; or (iv)
$100\times(1-1/(\sqrt{\mathrm{dm\_tol}})\%$ loss in S/N between DM
trials, i.e. $\sim 10.6\%$ loss for $\mathrm{dm\_tol}=1.25$. It is
neither of these. 

\textit{Caveat 1:} While the fourth interpretation is the most
intuitive, and what we effectively adopt below for our own
calculations, we re-iterate a subtle point made above. For
$\mathrm{dm\_tol}=1.25$ the $(i+1)^{\rm th}$ DM trial is chosen to be
at the DM where the response of the $(i)^{\rm th}$ DM trial reaches
$90.4\%$. The minimum of the scalloped response between the two trials
is never this low. With this caveat alone \textsc{heimdall} would
produce `too many' DM trials resulting in a better response than
expected/requested. However this is not the case as a second effect,
working in the opposite sense, is more impactful. \textit{Caveat 2:}
The DM smearing calculation used within \textsc{dedisp} uses an
approximation that applies best to narrow fractional bandwidths
centred around $1.4$~GHz. For frequencies far from $1.4$~GHz and/or
for large fractional bandwidths this assumption breaks down. One could
`correct' the DM trial calculations in \textsc{dedisp}, as used by
\textsc{heimdall}, but there are several objections to this: (i) the
calculation is not incorrect \textit{per se}, it is an approximation
which is correct under certain conditions; and (ii) it has been widely
deployed as is and these studies can not, and need not, be re-done if
the resulting detected samples are appropriately interpreted. Overall
we think the best thing to do is to clarify the interpretation of
dm\_tol.


\section{Results}
\begin{figure}
  \begin{center}
  \includegraphics[scale=0.32,angle=0,trim=10mm 20mm 10mm 20mm, clip]{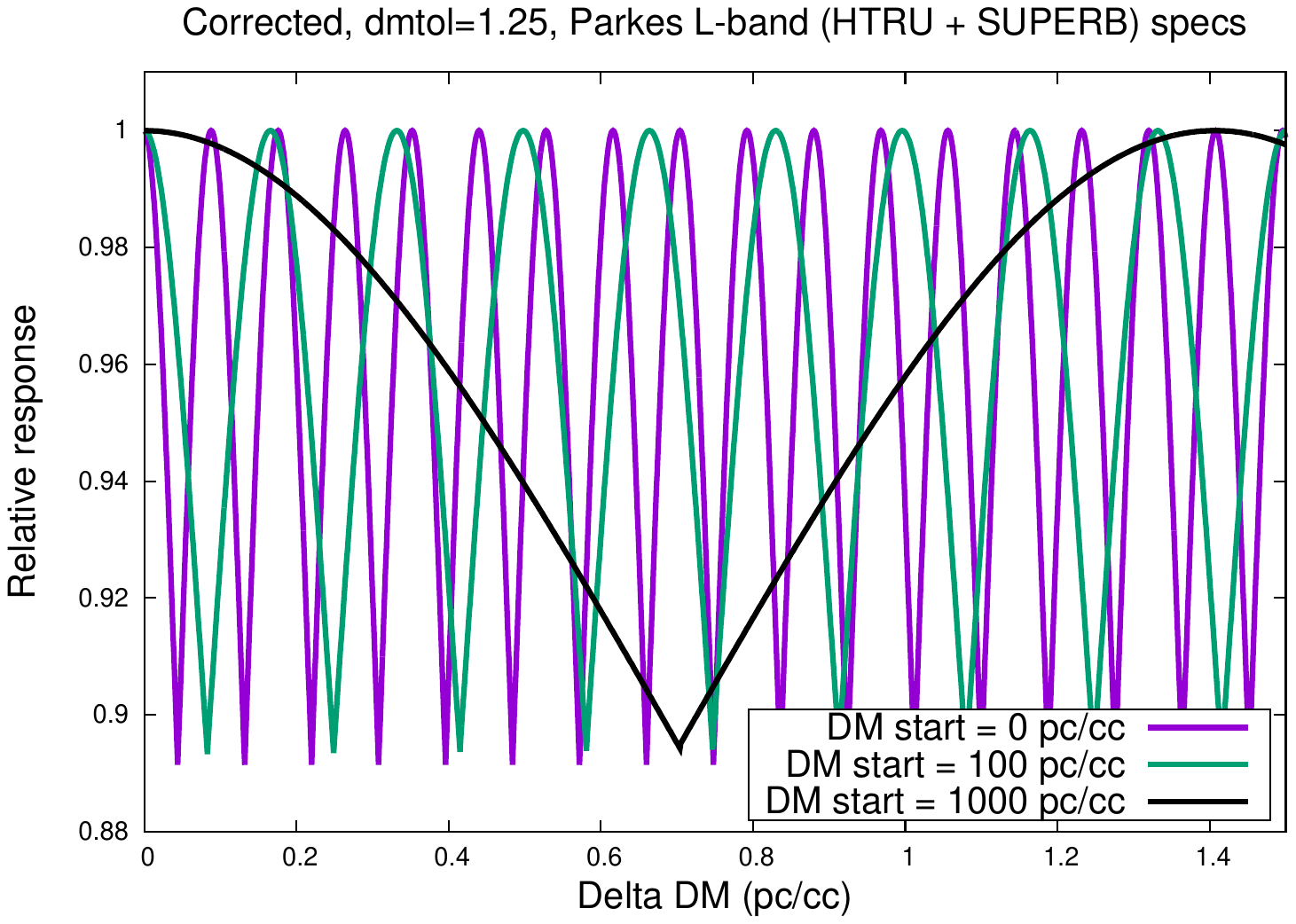}
  \includegraphics[scale=0.32,angle=0,trim=10mm 20mm 0mm 20mm, clip]{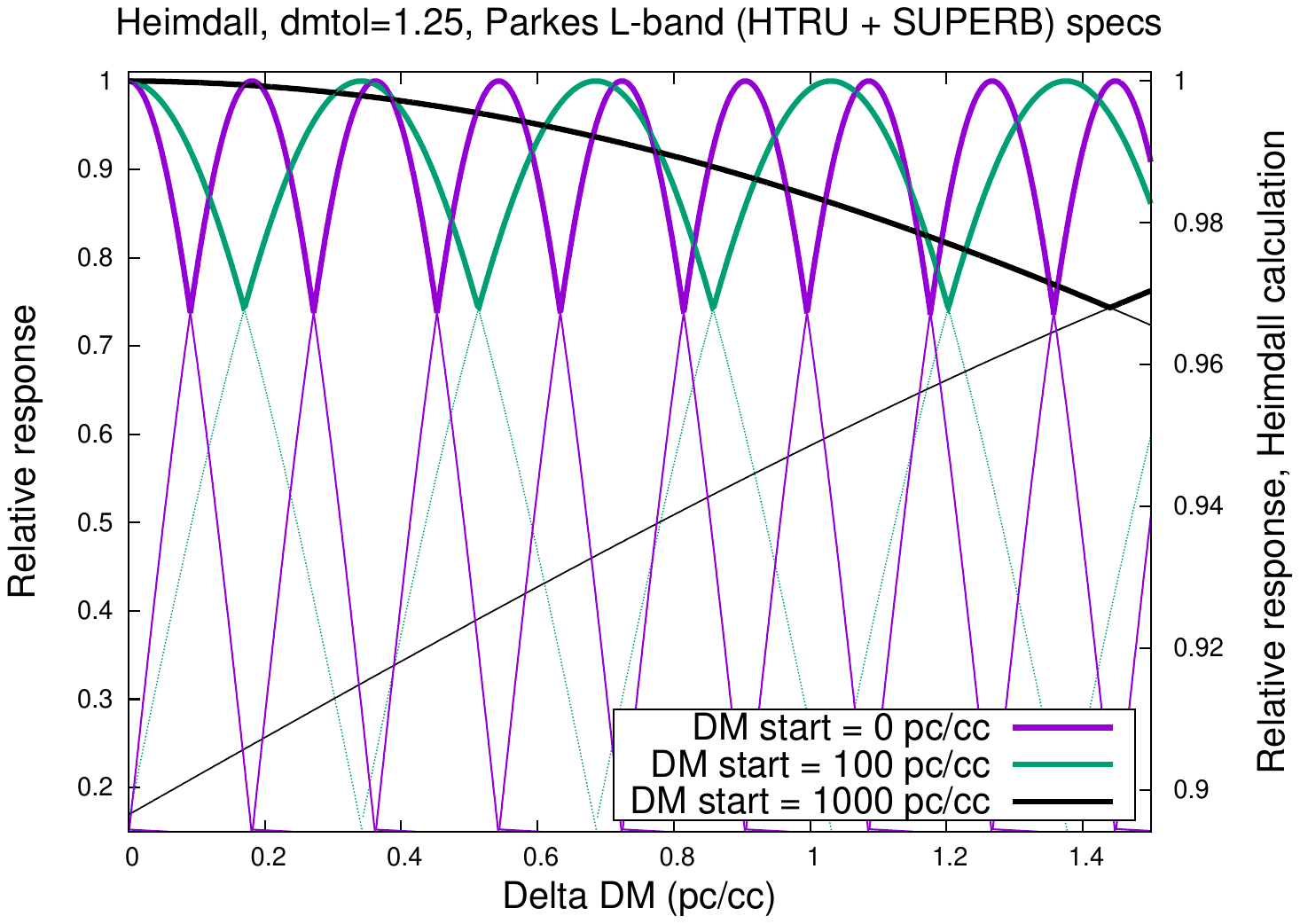}
  \caption{Left: the Parkes response for dm\_tol of $1.25$ shows
the minima of the sensitivity scallops reaching $1/\sqrt{1.25}=0.894$.
Responses for DM values of 0, 100 and 1000 pc/cc are shown for
illustration. Right: the response for DM trials determined
by \textsc{heimdall}. Here the right-most vertical axis shows the
relative response \textit{as calculated by} \textsc{heimdall}. 
The full scalloped response of the $(i)^{\rm th}$ trial is shown (lighter curves) all the
way to the $(i+1)^{\rm th}$ sample to illustrate how DM steps are
chosen. 
}\label{fig:response}
  \end{center}
\end{figure}

We interpret dm\_tol to mean that the minima of the scalloped DM
response should be at $1/\sqrt{\mathrm{dm\_tol}}$ of the maximum
response. We calculate DM trials for this interpretation and
illustrate in Fig~\ref{fig:response} the response for dm\_tol of
$1.25$ for specifications matching some legacy Parkes
surveys~\citep{htru,superb}. For these Parkes surveys there is a
simple linear relation between $x$, the desired dm\_tol input, and
$y$, the effective dm\_tol provided by the algorithm. The relation is:
$y=m(x-1)+c$ where $m=3.37(4)$ and $c=1.015(5)$. Thus if one desires
to know what dm\_tol to input into \textsc{heimdall} for a minimum
response of $75\%$ (say) it is given by: $x=1+(y-c)/m$, where
$y=1/(0.75)^2=1.77\dot{7}$, giving $x=1.226$. In this scenario this
means that if you \textit{want} a dm\_tol of $1.77\dot{7}$ you should
\textit{ask} for a dm\_tol of $1.226$. We also performed the same
calculations for a LOFAR high-band antenna observing setup, commonly
used at the Irish and other international LOFAR
stations~\citep{david_rrats}. Here the relation is steeper $m=4.38(9)$
and $c=1.03(1)$; to obtain a minimum $75\%$ response one would need to
ask for a dm\_tol of $1.171$. 

Understanding of selection effects such as those described above gives
us a more complete idea of the parameter space that was \textit{not}
searched, either partially or completely so. This is important for
correctly interpreting the populations detected by these
algorithms~\citep{hjq+24}. Additionally it can highlight areas where
improvements in the detection threshold can be made~\citep{qkb+23};
the brightness distributions of pulsars, FRBs etc. are such that even
very small improvements in the detection threshold can have a large
impact on detections~\citep{aje+25}. We freely provide the
code\footnote{\texttt{https://github.com/evanocathain/Useful\_FRB\_stuff/}}
used here so that the same calculations can be made for any
observational setup of interest to the reader.

\begin{acknowledgements}
EFK would like to thank Owen Johnson for encouragement to finish up
this note. 
\end{acknowledgements}

\begin{samepage}
\bibliographystyle{aasjournalv7}
\bibliography{rnaas_heimdall}
\end{samepage}

\end{document}